# Unconventional superconducting diode effects via antisymmetry and antisymmetry breaking

Chong Li[1,2,+], Yang-Yang Lyu[1,2,+], Wen-Cheng Yue[1,2,+] , Peiyuan Huang[1], Haojie Li[1], Tianyu Li[1], Chen-Guang Wang[1,2], Zixiong Yuan[1], Ying Dong[3], Xiaoyu Ma[4], Xuecou Tu[1], Tao Tao[5], Sining Dong[1,5,*], Liang He[5], Xiaoqing Jia[1], Guozhu Sun[1,2], Lin Kang[1], Huabing Wang[1,2,*], Francois M. Peeters[6], Milorad V. Milošević[6,*], Peiheng Wu[1,2] and Yong-Lei Wang[1,2,5,*]

[1]*Research Institute of Superconductor Electronics, School of Electronic Science and Engineering, Nanjing University, Nanjing, 210023, China*

[2]*Purple Mountain Laboratories, Nanjing, 211111, China*

[3]*College of Metrology & Measurement Engineering, China Jiliang University, Hangzhou, 310018, China*

[4]*Microsoft, One Microsoft Way, Redmond, WA 98052, USA*

[5]*State Key Laboratory of Spintronics Devices and Technologies, Nanjing University, Suzhou, 215163, China*

[6]*Department of Physics, University of Antwerp, Groenenborgerlaan 171, B-2020 Antwerp, Belgium*

+ Authors contribute equally

* Correspondence to: sndong@nju.edu.cn; hbwang@nju.edu.cn; milorad.milosevic@uantwerpen.be; yongleiwang@nju.edu.cn

## Abstract

**Symmetry-breaking plays a pivotal role in unlocking intriguing properties and functionalities in material systems. For example, the breaking of spatial and temporal symmetries leads to a fascinating phenomenon of superconducting diode effect. However, generating and precisely controlling the superconducting diode effect poses significant challenges. Here, we take a novel route with deliberate manipulation of magnetic charge**

potentials to realize unconventional superconducting flux-quantum diode effects. We achieve this through suitably tailored nanoengineered arrays of nanobar magnets on top of a superconducting thin film. We demonstrate the vital roles of inversion antisymmetry and its breaking in evoking unconventional superconducting effects—a magnetically symmetric diode effect and an odd-parity magnetotransport effect. These effects are non-volatilely controllable through in-situ magnetization switching of the nanobar magnets. Our findings promote the use of antisymmetry (breaking) for initiating unconventional superconducting properties, paving the way for exciting prospects and innovative functionalities in superconducting electronics.

Generally, the charge transport effect in a material exhibits even symmetry with respect to the externally applied electric current and magnetic field.[1] For instance, the resistance typically remains unchanged between reversed currents as well as between inverted magnetic fields. Recently, the superconducting diode effect with one-way supercurrent has gathered significant attention due to its potential for dissipationless electronics.[2-20] Such a diode shows zero resistance when current flows in one direction while exhibiting normal resistance when the current is reversed. Its manifestation typically relies on the simultaneous breaking of space inversion and time reversal symmetry.[2-15] Notably, the polarity of the diode can be reversed by inverting the applied magnetic field.[2-15] This phenomenon is generally described by the magnetochiral anisotropy, which incorporates a nonlinear resistance term ($\gamma \boldsymbol{B} \cdot \boldsymbol{I}$) dependent on both the applied electric current $I$ and the magnetic field $B$ ($\gamma$ denotes the magnetochiral anisotropy coefficient).[2-6] Consequently, a superconducting diode often exhibits a simultaneous nonreciprocal effect in response to both electric currents and magnetic fields. The intricate connections between the superconducting nonreciprocal charge transport and the symmetry breaking remains ambiguous, limiting the deliberate control of superconducting diodes.

On the other hand, a special type of superconducting diodes, relying on ratchet motion of superconducting flux-quanta, have been extensively investigated for years. [21-35] In type-II superconductors, magnetic flux penetrates the material in the form of individual flux-quanta (also known as vortices). The motion of such flux-quanta directly impacts the electromagnetic properties of superconductors.[36] Introducing symmetry-breaking in the pinning potentials leads to one-way motion of flux-quanta, thus presenting a flux-quantum diode. [21-34] However, akin to the other superconducting diodes, vast majority of previously reported flux-quantum diodes also display asymmetric responses with respect to both $I$ and $B$, [21-30] consistent with the description by magnetochiral anisotropy. [2-6] So far, whether magnetochiral-anisotropy plays a center role in superconducting diodes is still an open question.

In this study, we demonstrate, both experimentally and numerically, the inducing and manipulating two types of *unconventional* superconducting diode effects that cannot be described by magnetochiral anisotropy. We showcase a magnetically symmetric superconducting diode with a substantial diode efficiency up to 20%, as well as an electrically symmetric superconducting odd-parity magnetotransport effect, presenting a unique superconducting diode driven by magnetic fields instead of driven by electric currents. We achieve the first unconventional diode effect by introducing spatial inversion antisymmetic magnetic potential in the system, and then we break the antisymmetry to realize the later one. These unique ovel diode effects are produced in sophisticatedly designed hybrid devices, containing arrays of nanobar magnets on top of a superconducting thin film, as illustrated in Figures 1a.

Our device consists of a microbridge (50 nm thick and 50 μm wide) of amorphous superconducting $Mo_{0.79}Ge_{0.21}$ (MoGe) (Figure 1b), known for its weak intrinsic pinning, thus enabling the access of emergent effects from artificially introduced potential landscapes. [37,38] The superconducting critical temperature of our microstripe is 7.15 K (Supplemental Figure S1). The microbridge comprises four sections (S1-S4), each with square arrays of nanomagnets with different sizes and geometries (Figures 1c-1f) to demonstrate various superconducting effects. The detailed device fabrication procedures and parameters can be found in the Supporting Information. Each nanobar magnet can be considered as a pair of magnetic charges, one positive and one negative, respectively repelling and attracting the flux-quantum (inset of Figure 1l). [34, 39] The magnetic charge patterns of the nanomagnet arrays are directly observed through magnetic force microscopy (MFM) imaging, as shown in Figures 1g-1j.

Firstly, we present the generation of a distinctive superconducting diode that cannot be described by magnetochiral anisotropy. Figures 1k-1n display the $B$-dependent critical currents $I_c$ under positive (purple curve) and negative (green curve) currents from microbridge sections S1-S4, respectively. The $I_c$ curves are extracted from the $B$- and $I$-dependent dissipation voltage maps (Supplemental Figures S2). Clear peaks appear in all curves at the matching magnetic field $B_m$ = 114 Gs, corresponding to the flux-quantum density of one flux-quantum per nanomagnet. This indicates strong magnetic pinning provided by the nanomagnet arrays to the flux-quanta. In Figure 1k, the $I_c$ curves for microbridge S1 are symmetric to both $I$ and $B$. This results from the length of the rectangular nanomagnets in microbridge S1, matching the square lattice constant of 300 nm (Figure 1c), thereby generating a symmetric magnetic charge pattern, as depicted in Figures 2a and 2e. To break the space inversion symmetry, we elongate the nanomagnets in microbridge S2 (Figure 1d), where the nanomagnet length of 400 nm is longer than the square lattice constant.

This alteration breaks the space inversion symmetry (Figures 2b and 2f). Consequently, a significant discrepancy between the $I_c$ curves for positive and negative currents is observed (Figure 1l), leading to a flux-quantum diode effect. This current-driven nonreciprocal effect arises directly from the breaking of space inversion symmetry in the pinning potential [21-33, 39-42]. In our case, the symmetry-breaking originates from the asymmetric distribution of positive and negative magnetic charges (Figure 2f). The observed flux-quantum diode demonstrates a substantial superconducting diode efficiency ($[|I_c(+)-I_c(-)|]/[I_c(+)+I_c(-)]$) of up to 20% over a wide magnetic field range (Supplemental Figures S3).

Unlike traditional superconducting diodes, which typically display simultaneous nonreciprocal transport in response to both $I$ and $B$, [2-15, 21-30, 43, 44] the superconducting diode shown in Figure 1l exhibits a symmetric response to magnetic field $\boldsymbol{B}$, contrary to the principle of magnetochiral anisotropy. To unveil the underlining mechanism behind this unconventional superconducting diode effect, we analyze the symmetry of magnetic charges and their interaction with flux-quanta. In addition to the symmetry breaking (Figure 2f), we find the magnetic charge pattern also display inversion antisymmetry (Figure 2j). The antisymmetry plays a pivotal role in the magnetically symmetric transport properties within our devices, as elaborated below.

The inversion antisymmetry is described by (Figure 2j):

$$P(-r) = -P(r), \qquad (1)$$

where $P \rightarrow -P$ represents the inversion of magnetic charge potentials (positive to negative and vice versa), and $r \rightarrow -r$ denotes the space inversion operation. Furthermore, the interactions between magnetic charges and flux-quanta are opposite to those between identical magnetic charges and antiflux-quanta, as shown in the inset of Figure 1l. This relationship is denoted by:

$$P(-B) = -P(B), \qquad (2)$$

where $P(B)$ and $P(-B)$ represent the magnetic pinning potentials from the same charge pattern for flux-quanta and antiflux-quanta, respectively. Additionally, considering a given pinning potential $P$, the following relationships hold (refer to Supplemental Figures S4, S5 for details):

$$\begin{cases} V(-B, -I, P) = -V(B, I, P), & (3) \\ V[B, -I, P(-r)] = -V[B, I, P(r)], & (4) \end{cases}$$

where $V$ represents the dissipation voltage, directly measuring the motion of (anti)flux-quanta. Due to the forbidden of negative resistance, a sign reversal of $V$ is necessary whenever the polarity of the applied current $I$ is changed. Formula (3) states that the energy dissipation of flux-quanta while moving across a specific pinning potential equals that of antiflux-quanta (driven by reversed current) within the same pinning potential (see Supplemental Figure S4). Formula (4) indicates that the motion of flux-quanta across a pinning potential, under the space inversion operation, is equivalent to their motion under a reversed driving current across the original potential (Supplemental Figure S5). Combining formulas (1)-(4) yields:

$$V(-B) = V(B).$$

Hence, this suggests that space inversion antisymmetry leads to symmetric transport responses against the out-of-plane magnetic field $\boldsymbol{B}$, thus giving rise to a unique superconducting diode. This unconventional diode effect can be accurately replicated by our molecular dynamic (MD) simulations shown in Figure 2n and visually demonstrated in Video 1, Supporting Information. Snapshots captured from the video are presented in Supplemental Figure S6. Notably, the flux-quanta move only for one polarity of the current, clearly revealing a current-driven diode effect.

The above analysis highlights the significant role of antisymmetry in achieving a magnetically symmetric diode effect. To explore the emergent effect from antisymmetry-breaking, we design a distinctive nanomagnet array using asymmetric nanobars (Figure 1e). Both MFM imaging (Figure 1i) and micromagnetic simulation (inset of Figure 1i) reveal differences in the spatial spread and stray field strength of magnetic charges at the two ends of the asymmetric magnetic nanobar. The narrower end exhibits a more widely spread magnetic charge compared to the wider end, as depicted in inset of Figure 1n. Since the positive and negative charges from the same nanomagnet produce equal total flux, the focused charge generates a more localized yet stronger pinning force in contrast to the widely spread charge (inset of Figure 1n). These mismatches in magnetic potentials between positive and negative charges break the inversion antisymmetry of pinning potentials, signified by $P(-r) \neq -P(r)$ (illustrated in Figure 2k).

Both our experiment (Figure 1m) and MD simulation (Figure 2o) reveal an asymmetric response to the magnetic field. The $I_c$ at the negative fields (within the matching magnetic field) surpasses that at positive fields. This discrepancy gives rise to an odd-parity component of $I_c$, represented as $I_c^{odd} = [|I_c(B)| - |I_c(-B)|]/2$, showcasing a superconducting odd-parity magnetotransport effect.

Odd-parity magnetotransport is a rare phenomenon, observed in exceptional cases where time reversal symmetry is broken within a few material systems. [45-49] This uncommon field-driven nonreciprocal effect is typically subtle. [46-49] A recent study highlighted a substantial odd-parity magnetoresistance in an InAs quantum well,[45] emphasizing this intriguing effect in non-superconducting materials. While in superconductors, the odd-parity magnetotransport effect has been numerically explored in Josephson junctions subjected to a local inhomogeneous magnetic field, [50] experimental demonstrations of such unconventional superconducting effect are scarce. Our work presents an effective method for experimentally triggering this unconventional superconducting effect through antisymmetry-breaking.

Although asymmetric magnetic field responses have been widely demonstrated in superconducting diodes, [2-15, 21-30] satisfying magnetochiral anisotropy, these diodes exhibit a simultaneous current-driven nonreciprocal effect known as the regular diode effect. However, our observed superconducting odd-parity magnetotransport effect displays symmetric responses regarding to electric currents. As shown in Figures 1m and 2o, the $I_c$ curves under positive and negative currents overlap. Hence, this electrically symmetric yet magnetically asymmetric superconducting effect also rules out the mechanism of magnetochiral anisotropy. This sets apart the superconducting odd-parity magnetotransport effect induced by antisymmetry breaking from any previously reported superconducting diode effects satisfying magnetochiral anisotropy, [2-15, 21-30] typically induced by standard spatial-temporal symmetry breaking. The electrically symmetric response observed in microbridge S3 occurs because the length of the asymmetric nanobar matches the square lattice constant of 300 nm, hence the antisymmetric magnetic charge pattern preserves the inversion symmetry (Figure 2g).

The odd-parity magnetotransport effect is visualized in MD simulation Video 2 (Supporting Information). Snapshots captured from the video are presented in Figure S7. Notably, the flux-quanta move solely for one polarity of the applied magnetic fields. This behavior resembles that of a current-driven flux-quantum diode, where flux-quanta move exclusively for only one polarity of applied current, suggesting a magnetic-field-driven diode effect.

We can merge both the current-driven diode effect and the magnetic-field-driven diode effect into one device by simultaneously breaking inversion symmetry and antisymmetry. This can be achieved by extending the length of asymmetric nanomagnets, as depicted in Figure 1f. The corresponding magnetic charge pattern (shown in Figures 1j and 2d) reveals simultaneous symmetry-breaking (Figure 2h) and antisymmetry-breaking (Figure 2l). As anticipated, the $I_c$ curves (Figure 1n in the experiment and Figure 2p in the simulation) display asymmetry concerning both $I$ and $B$, resembling the behavior of traditional superconducting diode.

However, different from the traditional superconducting diodes where the diode polarity can be inverted by changing the direction of the magnetic field, in Figure 1n, the diode polarity remains the same between positive and negative magnetic fields despite clear asymmetric magnetic field dependence. This discrepancy arises because reversing the current does not alter the pinning strengths determined by the magnetic charges, although it does reverse the flow direction of flux-quanta.

To further demonstrate the deliberate unconventional superconducting diode effects, we showcase the effects of switching current and magnetic field in Figure 3. The red curves display the resistance switching effect upon alternatively reversing electric currents, while the blue curves display the switching effect upon alternatively reversing magnetic fields. The results can be categorized into four groups based on the parity operation of inversion symmetry and antisymmetry: 1) No diode switching effects under simultaneously preserved inversion symmetry and antisymmetry; 2) Pure current-driven diode switching occurs when inversion symmetry is violated while antisymmetry is preserved; 3) Pure magnetic-field-driven diode switching occurs when antisymmetry is violated while inversion symmetry is preserved; and 4) Both current- and magnetic-field-driven diode switching occurs when both inversion symmetry and antisymmetry are violated. These results suggest that magnetic symmetry-breaking solely governs the current-driven diode effect, while the antisymmetry-breaking solely governs the magnetic-field-driven diode (or odd-parity magnetotransport) effect. Notably, all these unconventional superconducting effects cannot be described by a magnetochiral-anisotropy picture.

In traditional superconducting diodes, [2-15, 21-30] the reversal of the nonreciprocal effect against current (or magnetic field) could be achieved by changing the direction of magnetic field (or current), as required by magnetochiral anisotropy. However, as mentioned above, all nonreciprocal effects in our devices—both electrically and magnetically—violate the principle of magnetochiral anisotropy. Consequently, their polarities cannot be altered by switching currents or magnetic fields.

However, we can reverse diode polarity through a novel and nonvolatile method by inverting the magnetization states of nanomagnets. This inversion can be achieved through in-situ applied in-plane magnetic fields. As depicted in Figures 4a-4d, when the magnetization state of nanomagnets is inverted, the polarities of magnetic charges switch, while their distributions and strengths remain unchanged. This inverts the orientations of symmetry-breaking and antisymmetry-breaking, consequently reversing the polarities for all electric and magnetic nonreciprocal effects (as shown in Figures 4e-4h). The reversed diode effects are also replicated by our molecular dynamic (MD) simulations (Figure S8, Supporting Information).

We have successfully demonstrated how the introduction and subsequent breaking of space inversion antisymmetry in magnetic charge potentials can unlock unconventional properties within superconductors. Specifically, by incorporating space inversion antisymmetry, we have facilitated the emergence of a magnetically symmetric superconducting diode. More intriguingly, intentionally disrupting this antisymmetry has proven to be an effective strategy for activating the odd-parity magnetotransport effect, leading to a unique magnetic-field-driven superconducting diode. A schematic phase diagram of major superconducting responses in regard to symmetry (breaking) and antisymmetry (breaking) is displayed in Figure S9, Supporting Information. To be noted, our results unambiguously rule out the role of magnetochiral anisotropy in our superconducting diodes.

These findings serve as a conceptual proof-of-principle and invite further exploration of emergent properties associated with inversion antisymmetry and its breaking, holding potential for application across diverse material systems hosting (quasi)particles beyond superconducting flux-quanta. For instance, this can be extended to magnetic colloids, [51,52] skyrmions, [53] and magnetic vortices, [54] where the polarities of these particles can also be controlled by magnetic fields. Moreover, by extending this manipulation to electric (quasi)particles, the introduction of unique electric potentials with antisymmetry or its breaking could lead to innovative transport phenomena. This could be realized by fabricating nanopatterned gating electrodes in two-dimensional materials like graphene,[55] where the charge polarities (electrons and holes) can be controlled using an electric field. Hence, the future exploratory use of inversion antisymmetry and its breaking holds great potential for discovering new principles and functionalities, paving the way for innovative applications.

**Supporting Information**

Supporting Information is available from Online or from the author.

Videos S1 and S2 showing results of MD simulations for unconventional superconducting vortex diode effects with inversion antisymmetry and antisymmetry breaking (MP4, MP4).

Methods for sample fabrication, experiments, molecular dynamic simulations and micromagnetic simulation, Supplemental Figures S1-S9.

**Notes**

The authors declare no competing financial interest.

**Acknowledgements**

This work is supported by the National Natural Science Foundation of China (62288101, 62274086 and 12204434), the National Key R&D Program of China (2021YFA0718802 and 2023YFF0718400), the Fundamental Research Funds for the Central Universities, and Jiangsu Key Laboratory of Advanced Techniques for Manipulating Electromagnetic Waves.

**Figure 1**

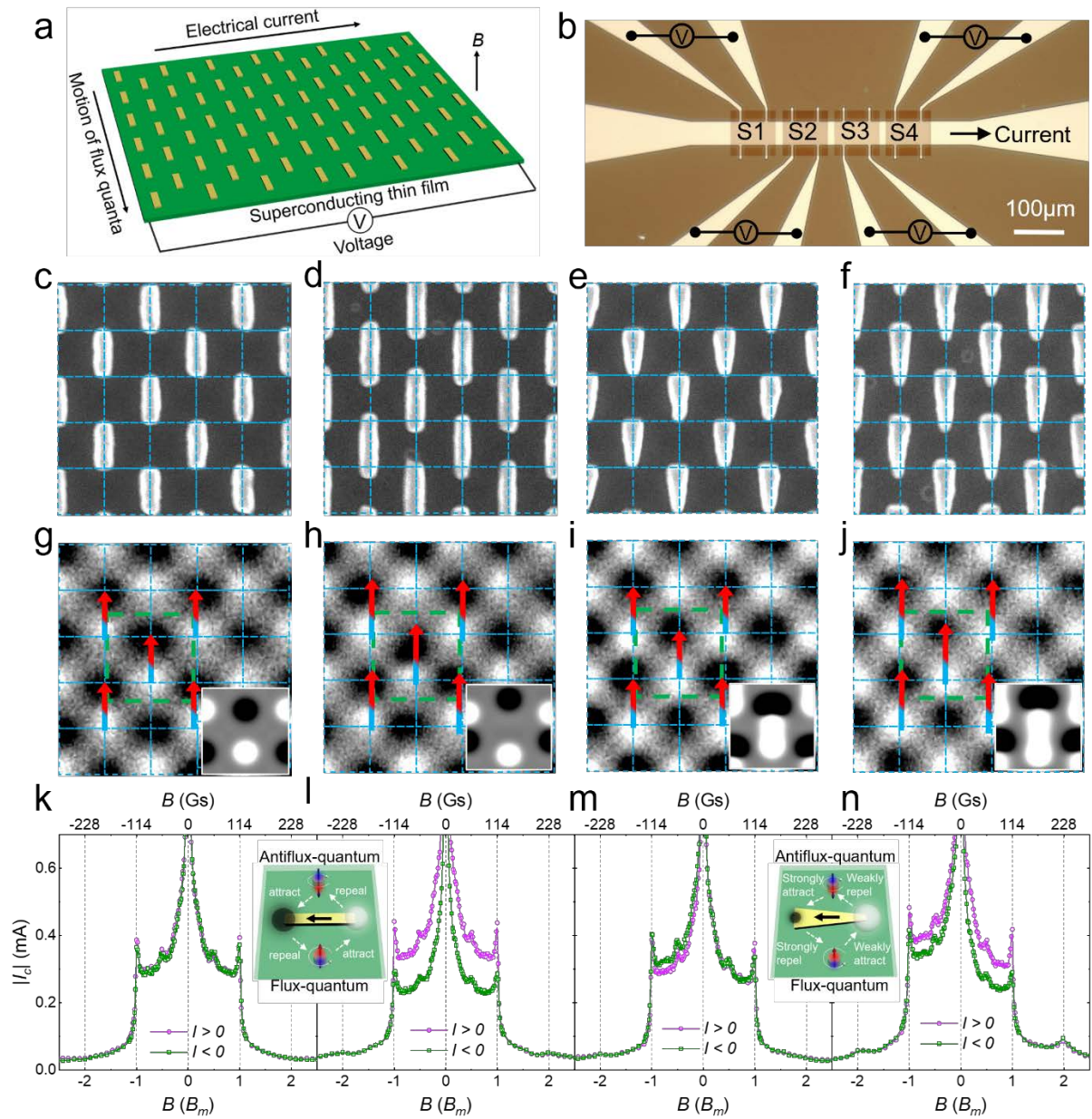

**Figure 1. Unconventional flux-quantum diode effects. a**, Schematic diagram illustrating an array of nanomagnets on top a superconducting film. **b**, Optical image of a superconducting MoGe microbridge with four sections, patterned with square arrays of nanomagnets shown in (c-f), respectively. Scale bar, 100 µm. **c-f**, Scanning electron microscopy (SEM) images showing square arrays of nanomagnets structure. The square lattice constant is 300nm. **g-j**, MFM images of the nanomagnets arrays in (c)-(f). The black/white spots reveal positive/negative magnetic charges, while the arrows indicate the magnetizations of the single-domain nanomagnets. The inset displays the images obtained from micromagnetic simulations. **k-n**, Comparisons of critical current curves for positive and negative currents from the four sections S1-S4, respectively. The insets display the interactions between a nanomagnet and a flux-quantum (bottom) and an antiflux-quantum(top).

**Figure 2**

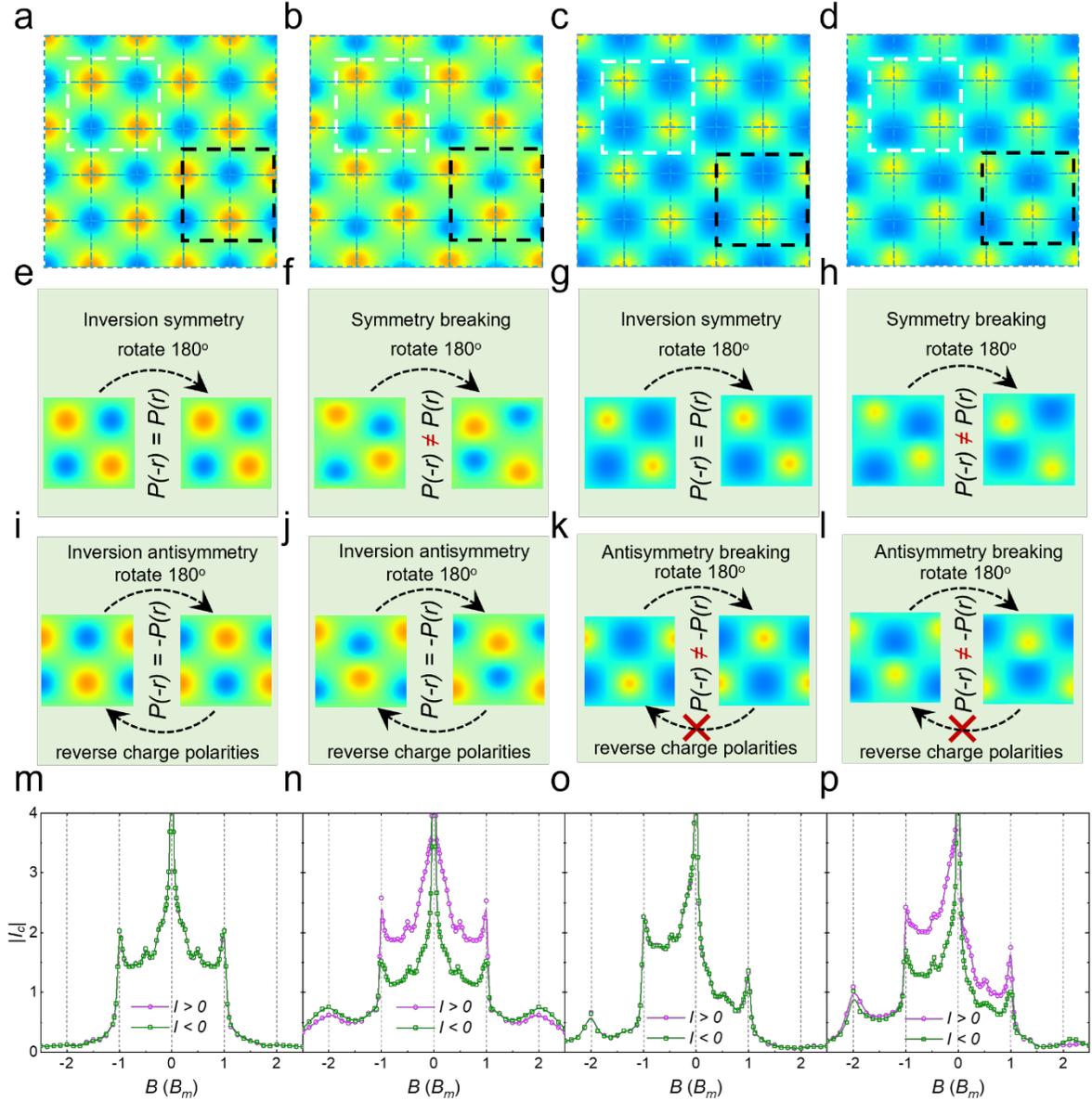

**Figure 2. Theoretical models and numerical simulations of unconventional diode effects. a-d**, models of magnetic charge potentials corresponding to Figure 1g-j, respectively. The blue/red spots represent attractive/repulsive pinning potentials to flux-quanta. White and black boxes highlight the unit cells of the same size as described in (e-h) and (i-l), respectively. **e-h**, Illustration of space inversion symmetry and its breaking. **i-l**, Illustration of space inversion antisymmetry and its breaking. **m-p**, MD simulations of $|I_c| \sim B$ curves using the potentials in (a-d), respectively.

**Figure 3**

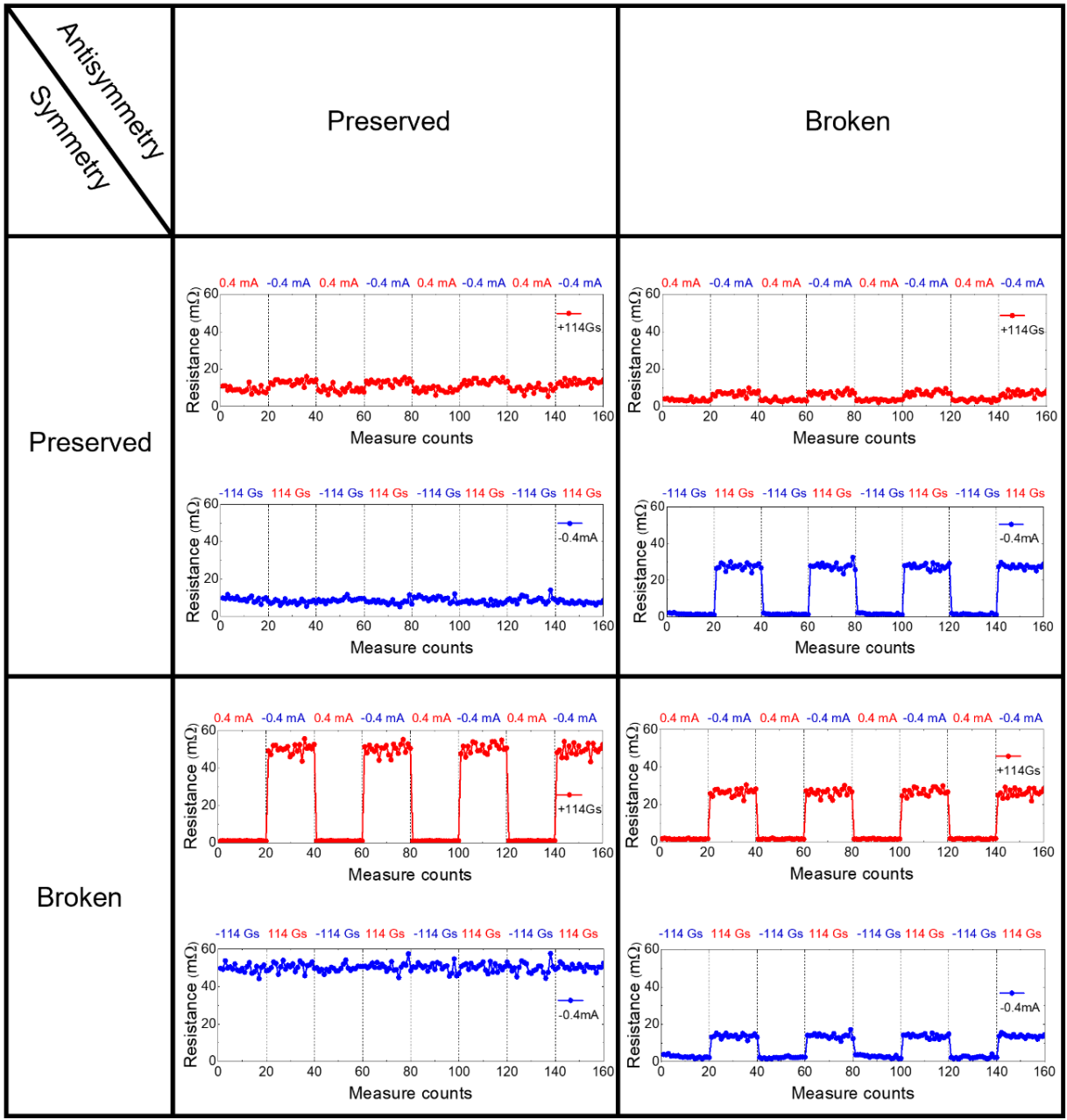

**Figure 3. Switchable superconductivity.** The red and blue curves display the resistance switching by alternatively reversing electric currents (red) and reversing magnetic fields (blue).

**Figure 4**

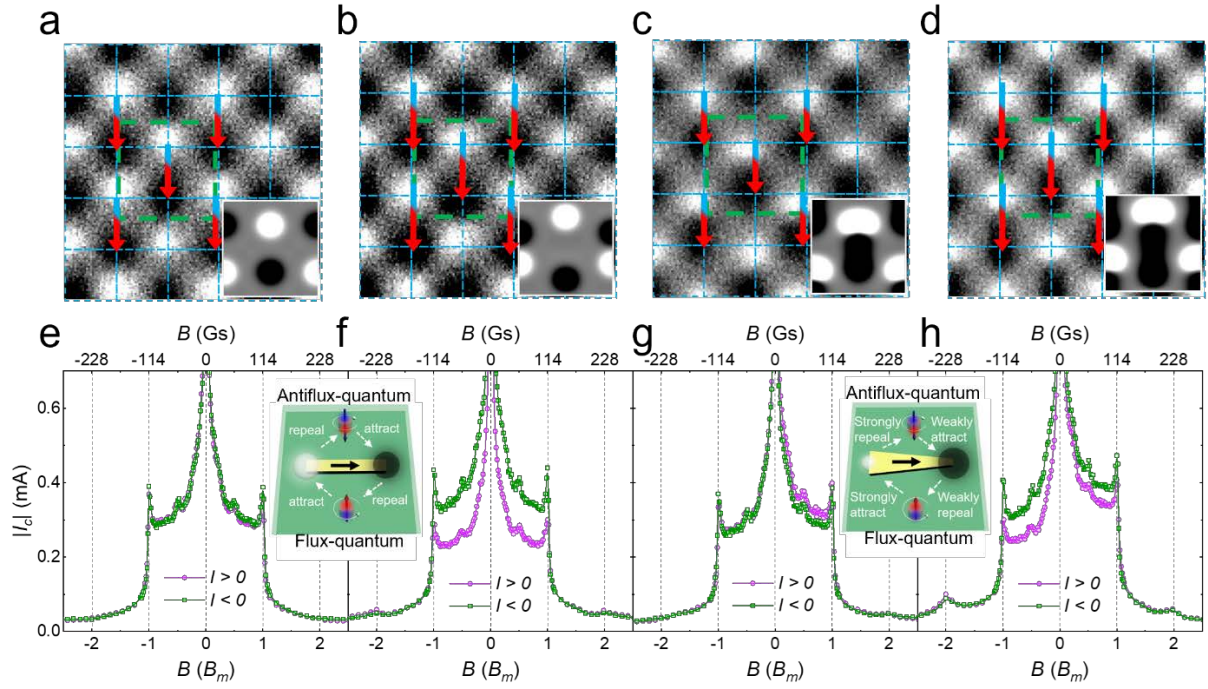

**Figure 4. Reversing diode polarities.** a-d, MFM images of the nanomagnets arrays (Figures 1c-1f) under reversed magnetization. Insets are images from micromagnetic simulation. **e-h**, Experimental $I_c \sim B$ curves corresponding to (a-d), respectively. The insets illustrate the reversed interactions between nanomagnets and (anti)flux-quantum.